\documentclass[12pt]{article}
 
\usepackage{latexsym,amssymb,epsfig}
\newcommand{\eqref}[1]{(\ref{#1})}
\def\Z{\mathbb{Z}}
\def\fnote#1#2{\begingroup\def\thefootnote{#1}\footnote{#2}
     \addtocounter{footnote}{-1}\endgroup}
\newcommand{\Rep}[1]{\ensuremath{\mathbf{#1}}}
\newcommand{\barRep}[1]{\ensuremath{\overline{\Rep{#1}}}}
\newcommand{\Vt}{{\ensuremath{\widetilde{V}}}}
\newcommand{\Xt}{{\ensuremath{\widetilde{X}}}}

\DeclareFontFamily{OT1}{rsfs10}{}
\DeclareFontShape{OT1}{rsfs10}{m}{n}{ <-> rsfs10 }{}
\DeclareMathAlphabet{\mathscript}{OT1}{rsfs10}{m}{n}

\begin{document}
 
\begin{titlepage}
 
\vspace{-2cm}

\title{
   \hfill{\normalsize  UPR-1103-T} \\[1em]
   {\LARGE A Heterotic Standard Model\\
\author{Volker Braun$^{1,2}$, Yang-Hui He$^{1}$, Burt A.~Ovrut$^{1}$, 
        and Tony Pantev$^{2}$
        \fnote{~}{vbraun,
        yanghe, ovrut@physics.upenn.edu;
        tpantev@math.upenn.edu}\\[0.5cm]
   {\normalsize $^1$
        Department of Physics,}
   {\normalsize $^2$ 
        Department of Mathematics}\\
   {\normalsize University of Pennsylvania} \\
        {\normalsize Philadelphia, PA 19104--6395, USA}}
}
\date{}
}

\maketitle
 
\begin{abstract}
  
  Within the context of the $E_8 \times E_8$ heterotic superstring
  compactified on a smooth Calabi-Yau threefold with an $SU(4)$ gauge
  instanton, we show the existence of simple, realistic $N=1$
  supersymmetric vacua that are compatible with low energy particle
  physics. The observable sector of these vacua has gauge group
  $SU(3)_{C} \times SU(2)_{L} \times U(1)_{Y} \times U(1)_{B-L}$,
  three families of quarks and leptons, each with an additional {\it
  right-handed neutrino}, {\it two} Higgs-Higgs conjugate pairs, a
  small number of uncharged moduli and {\it no exotic matter}.  The
  hidden sector contains non-Abelian gauge fields and moduli. In the
  strong coupling case there is {\it no exotic matter}, whereas for
  weak coupling there are a {\it small number} of additional matter
  multiplets in the hidden sector. The construction exploits a 
  mechanism for ``splitting'' multiplets. The minimal nature and rarity of
  these vacua suggest the possible theoretical and experimental
  relevance of spontaneously broken $U(1)_{B-L}$ gauge symmetry and
  two Higgs-Higgs conjugate pairs. The $U(1)_{B-L}$ symmetry helps to 
  naturally suppress the rate of nucleon decay.

\end{abstract}

\thispagestyle{empty}
 
\end{titlepage}

The discovery of non-vanishing neutrino masses indicates that, 
in supersymmetric theories without exotic multiplets, a
right-handed neutrino must be added to each family of quarks and
leptons~\cite{neutrino}. It is well known that this augmented 
family fits exactly into
the $\Rep{16}$ spin representation of $Spin(10)$, making this group
very compelling from the point of view of grand unification and string
theory. Within the context of $N=1$ supersymmetric $E_8 \times E_8$
heterotic string vacua, a $Spin(10)$ group can arise from the
spontaneous breaking of the observable sector $E_8$ group by an
$SU(4)$ gauge instanton on an internal Calabi-Yau
threefold~\cite{WittenNew}. The $Spin(10)$ group is then broken by a
Wilson line to a gauge group containing $SU(3)_{C} \times SU(2)_{L}
\times U(1)_{Y}$ as a factor~\cite{wilson}. To achieve this, the
Calabi-Yau manifold must have, minimally, a fundamental group $\Z_3
\times \Z_3$.

Until now, such vacua could not be constructed since a) Calabi-Yau
threefolds with fundamental group $\Z_3 \times \Z_3$ were not known
and b) it was unknown how to find $SU(4)$ gauge instantons on such
manifolds. Recently, the first problem was rectified in~\cite{volker}.
We have now solved the second problem, exhibiting a large class of
$SU(4)$ gauge instantons on the Calabi-Yau manifolds presented
in~\cite{volker}. Generalizing the results in~\cite{extension,z2},
these instantons are obtained as connections on stable, holomorphic
vector bundles with structure group $SU(4)$. The technical details
will be given elsewhere~\cite{future}. In addition to these considerations, 
we also use a natural method for ``splitting'' multiplets 
that was introduced for general bundles in~\cite{z2}. In this paper, 
we present the results of our search for realistic vacua in this context.

The results are very encouraging. We find $N=1$ supersymmetric vacua
whose minimal observable sector, for both the weakly and strongly
coupled heterotic string, has the following properties.

\begin{itemize}

\item {Observable Sector: Weak and Strong Coupling} 

\begin{enumerate}

\item Gauge group $SU(3)_{C} \times SU(2)_{L} \times U(1)_{Y} \times
U(1)_{B-L}$. 

\item {\it Three families} of quarks and leptons, each with a {\it
right-handed neutrino}. 

\item {\it Two} Higgs-Higgs conjugate pairs.

\item {\it Six} geometric moduli and a {\it small number} of vector
bundle moduli. 

\item {\it No exotic matter fields}.

\end{enumerate}

\end{itemize}

These are, to our knowledge, the first vacua in any string theory
context whose observable sector contains no exotic matter. We
emphasize that, although very similar to the supersymmetric standard
model, our observable sector differs in three significant ways. First,
there is an extra right-handed neutrino in each family. Closely
related to this is the appearance of an additional gauged $B-L$
symmetry. Finally, we find, not one, but two Higgs-Higgs conjugate
pairs.

The structure of the hidden sector depends on whether one is in the
weakly or strongly coupled regime of the heterotic string. In the
strongly coupled context, we find the following minimal hidden sector.

\begin{itemize}

\item{Hidden Sector: Strong Coupling}

\begin{enumerate}

\item Gauge group $E_7 \times U(6)$.

\item A {\it small number} of vector bundle moduli.

\item {\it No matter fields.}

\end{enumerate}

\end{itemize}

Again, note that this hidden sector has no exotic matter. Combining
this with the above, we have demonstrated, within the context of the
strongly coupled heterotic string, the existence of realistic vacua
containing no exotic matter fields. We emphasize that the hidden
sector gauge group $E_7 \times U(6)$ is sufficiently large to allow
acceptable supersymmetry breaking via condensation of its gauginos.

In the weakly coupled context, we find the following minimal hidden
sector. (This is also a valid vacuum in the strongly coupled case).

\begin{itemize}

\item{Hidden Sector: Weak Coupling}

\begin{enumerate}

\item Gauge group $Spin (12)$.

\item A {\it small number} of vector bundle moduli.

\item {\it Two} matter field multiplets in the $\Rep{12}$ of $Spin(12)$.

\end{enumerate}

\end{itemize}

Note that, in this case, there are a small number of exotic matter
multiplets in the hidden sector. Again, the hidden sector gauge group
$Spin(12)$ is sufficiently large to allow acceptable supersymmetry
breaking via gaugino condensation.

The vacua presented above are the result of an extensive search within
the wide context made precise in~\cite{future}. They appear to be the
minimal vacua, all others containing exotic matter fields, either in
the observable sector, the hidden sector, or both, usually with a
large number of Higgs-Higgs conjugate pairs. We have been unable to
find any vacuum in this context with only a single pair of Higgs-Higgs
conjugate fields. Furthermore, to our knowledge, phenomenological
vacua in all other string contexts
\cite{z2,chsw,greene,pheno-orb,pheno-II} have substantial amounts of
exotic matter, both in the observable and hidden sectors. For all these 
reasons, we refer to the class of vacua presented in this paper 
as a {\it heterotic standard model} and speculate that it may be of
phenomenological significance. In particular, it would seem to
motivate renewed interest, both theoretical and experimental, in its
characteristic properties; namely, 1) the physics of a $U(1)_{B-L}$
gauge symmetry spontaneously broken at, or above, the electroweak
scale and 2) the physics of two pairs of Higgs-Higgs conjugate fields,
particularly their experimental implications for flavor changing
neutral currents. It is immediately clear that the $B-L$ symmetry will
help to naturally suppress the rate of nucleon decay.  This
potentially resolves a long-standing problem in phenomenological
string vacua. At the least, our results go a long way toward
demonstrating that realistic particle physics can be the low energy
manifestation of the $E_8 \times E_8$ heterotic superstring, as
originally envisaged in~\cite{chsw,origin}.

We now specify, in more detail, the properties of the these minimal
vacua and indicate how they are determined. Following~\cite{chsw}, the
requisite Calabi-Yau threefold, $X$, is constructed as follows. We
begin by considering a simply connected Calabi-Yau threefold, $\Xt$,
which is an elliptic fibration over a rational elliptic surface,
$d\mathbb{P}_9$. In a six-dimensional region of moduli space, such
manifolds can be shown to admit a $\Z_3 \times \Z_3$ group action
which is fixed point free. It follows that
\begin{equation}
X=\frac{\Xt}{\Z_3 \times \Z_3}
\label{1}
\end{equation}
is a smooth Calabi-Yau threefold that is torus-fibered over a singular
$d\mathbb{P}_9$ and has non-trivial fundamental group
\begin{equation}
\pi_{1}(X)=\Z_3 \times \Z_3
\,,
\label{2}
\end{equation}
as desired. It was shown in~\cite{volker} that $X$ has 
\begin{equation}
h^{1,1}(X)=3 \,, \quad 
h^{2,1}(X)=3
\label{3}
\end{equation}
K{\"a}hler and complex structure moduli respectively. To our
knowledge, this is the only Calabi-Yau threefold with $\Z_3 \times
\Z_3$ fundamental group that has been constructed. We note~\cite{phil}
that the transpose of the configuration matrix~\cite{hubsch}
associated with $\Xt$ defines another simply connected Calabi-Yau
threefold.  Interestingly, this is precisely the manifold introduced
by Tian and Yau~\cite{TY} which, when quotiented by $\Z_3$, was used
to construct three generation heterotic string vacua within the
context of the standard gauge embedding.

We now construct a stable, holomorphic vector bundle, $V$, on $X$ with
structure group
\begin{equation}
G= SU(4)  
\label{4}
\end{equation}
contained in the $E_8$ of the observable sector. This bundle admits a
gauge connection satisfying the Hermitian Yang-Mills equations. The
connection spontaneously breaks the observable sector $E_8$ gauge
symmetry to
\begin{equation}
E_8 \longrightarrow Spin(10)
\,,
\label{5}
\end{equation}
as desired. We produce $V$ by building stable, holomorphic vector
bundles $\Vt$ with structure group $SU(4)$ over $\Xt$ that are
equivariant under the action of $\Z_3 \times \Z_3$. This is
accomplished by generalizing the method of ``bundle extensions''
introduced in~\cite{extension}. The bundle $V$ is then given as
\begin{equation}
V=\frac\Vt{\Z_3 \times \Z_3}
\,.
\label{6}
\end{equation}

Realistic particle physics phenomenology imposes additional
constraints on $\Vt$. To ensure that there are three generations of
quarks and leptons after quotienting out $\Z_3 \times \Z_3$ one must
require that
\begin{equation}
c_{3}\big( \Vt \big)=\pm 54
\,,
\label{7}
\end{equation}
where $c_{3}(\Vt)$ is the third Chern class of $\Vt$. Recall that with
respect to $SU(4) \times Spin(10)$ the adjoint representation of $E_8$
decomposes as
\begin{equation}
  \Rep{248}=
  \big( \Rep{1},\Rep{45} \big) \oplus 
  \big( \Rep{15},\Rep{1} \big) \oplus 
  \big( \Rep{4},\Rep{16} \big) \oplus 
  \big( \barRep{4}, \barRep{16} \big) \oplus 
  \big( \Rep{6},\Rep{10} \big)
\,.
\label{8}
\end{equation}
The number of $\barRep{16}$ zero modes is given by
$h^1\big(\Xt,\Vt^*\big)$~\cite{z2}. Therefore, if we demand that
there be no exotic matter fields arising from vector-like
$\barRep{16}-\Rep{16}$ pairs, $\Vt$ must be constrained so that
\begin{equation}
h^1\left( \Xt, \Vt^* \right)=0
\,.
\label{9}
\end{equation}
Similarly, the number of $\Rep{10}$ zero modes is
$h^1\big(\Xt,{\wedge}^{2}\Vt\big)$. However, since the Higgs fields
arise from the decomposition of the $\Rep{10}$, we must not set the
associated cohomology to zero. Rather, we restrict $\Vt$ so that
$h^1\big(\Xt,{\wedge}^{2}\Vt\big)$ is minimal, but non-vanishing.
Subject to all the constraints that $\Vt$ must satisfy, we find
\begin{equation}
h^1\left( \Xt,{\wedge}^{2}\Vt \right)=14
\,.
\label{10}
\end{equation}
Finally, for the gauge connection to satisfy the Hermitian Yang-Mills
equations the holomorphic bundle $\Vt$ must be stable. A complete
proof of the stability of $\Vt$ is technically very involved and has
not been carried out. However, there are a number of non-trivial
checks of stability that can be made. Specifically, stability
constrains the cohomology of $\Vt$ to satisfy
\begin{equation}
h^0\left( \Xt,\Vt \right)=0 \,, \quad 
h^0\left( \Xt,\Vt^* \right)=0 \,, \quad 
h^0\left(\Xt, \Vt \otimes \Vt^*\right)=1
\,.
\label{11}
\end{equation}
We have shown~\cite{future} that vector bundles $\Vt$ satisfying the
constraints eqs.~\eqref{7},~\eqref{9},~\eqref{10} and~\eqref{11}
indeed exist. Henceforth, we will restrict our discussion to such
bundles.

We now extend the observable sector bundle $V$ by adding a Wilson
line, $W$, with holonomy
\begin{equation}
\mathrm{Hol}(W)=\Z_3 \times \Z_3 \subset Spin(10)
\,.
\label{12}
\end{equation}
The associated gauge connection spontaneously breaks $Spin(10)$ as
\begin{equation}
Spin(10) \longrightarrow 
SU(3)_{C} \times 
SU(2)_{L} \times 
U(1)_{Y} \times 
U(1)_{B-L}
\,,
\label{13}
\end{equation}
where $SU(3)_{C} \times SU(2)_{L} \times U(1)_{Y}$ is the standard model gauge
group. Since $\Z_3 \times \Z_3$ is Abelian and 
rank($Spin(10))=5$, an additional rank one factor must appear.
For the chosen embedding of $\Z_3 \times \Z_3$, 
this is precisely the gauged $B-L$ symmetry.

As discussed in~\cite{z2}, the zero mode spectrum of $V \oplus W$ on
$X$ is determined as follows. Let $R$ be a representation of
$Spin(10)$, and denote the associated $\Vt$ bundle by $U_{R}(\Vt)$. Find
the representation of $\Z_3 \times \Z_3$ on $H^1\big( \Xt,U_{R}(\Vt)\big)$
and tensor this with the representation of the Wilson line on $R$. The
zero mode spectrum is then the invariant subspace under this joint
group action. Let us apply this to the case at hand.  First consider
the $\barRep{16}$ representation. It follows from eq.~\eqref{9} that
no such representations occur. Hence, no exotic $SU(3)_{C} \times
SU(2)_{L} \times U(1)_{Y} \times U(1)_{B-L}$ fields arising from
vector-like $\barRep{16}-\Rep{16}$ pairs appear in the spectrum, as
desired. Now examine the $\Rep{16}$ representation. The Atiyah-Singer
index theorem, eq.~\eqref{7} and~\eqref{9} imply that
\begin{equation}
h^1\left( \Xt,\Vt \right)=27
\,.
\label{14}
\end{equation}
We can calculate the $\Z_3 \times \Z_3$ representation on
$H^1\big(\Xt,\Vt\big)$ as well as the Wilson line action on
$\Rep{16}$. Tensoring these together, we find that the invariant
subspace consists of three families of quarks and leptons, each family
transforming as
\begin{equation}
\big(\Rep{3},   \Rep{2}, 1, 1 \big) \,,\quad
\big(\barRep{3},\Rep{1}, -4, -1 \big) \,,\quad
\big(\barRep{3},\Rep{1}, 2, -1 \big)
\label{15}
\end{equation}
and
\begin{equation}
\big(\Rep{1},\Rep{2}, -3, -3 \big) \,,\quad
\big(\Rep{1},\Rep{1}, 6, 3 \big) \,,\quad
\big(\Rep{1},\Rep{1}, 0, 3 \big)
\label{16}
\end{equation}
under $SU(3)_{C} \times SU(2)_{L} \times U(1)_{Y} \times U(1)_{B-L}$.
We have displayed the quantum numbers $3Y$ and $3(B-L)$ for
convenience. Note from eq.~\eqref{16} that each family contains a
right-handed neutrino, as desired.

Finally, consider the $\Rep{10}$ representation. Recall from
eq.~\eqref{10} that $h^1\big(\Xt,{\wedge}^{2}\Vt\big)=14$. We find
that the representations of the two generators of $\Z_3 \times \Z_3$
in $H^1\big(\Xt,{\wedge}^{2}\Vt\big)$ are given by the $14 \times 14$
matrices
\begin{equation}
\mathrm{diag}(
1,1,1,
\omega_1,\omega_1^2,\omega_1,\omega_1^2,
1,1,1,
\omega_1,\omega_1^2,\omega_1,\omega_1^2)
\label{17}
\end{equation}
and
\begin{equation}
\mathrm{diag}(
1,\omega_2,\omega_2^2,
1,1,\omega_2^2,\omega_2,1,
\omega_2,\omega_2^2,1,
1,\omega_2^2,\omega_2)
\label{18}
\end{equation}
respectively, where $\omega_1$ and $\omega_2$ are third roots of
unity. Furthermore, the Wilson line $W$ can be chosen so that
\begin{equation}
\Rep{10}=(\omega_1^2) \Rep{5} \oplus (\omega_1) \barRep{5}
\label{19}
\end{equation}
and
\begin{equation}
\Rep{10}= 
\Big(\Rep{2} \oplus (\omega_2^2)\Rep{3} \Big) \oplus 
\Big(\barRep{2} \oplus (\omega_2) \barRep{3} \Big)
\label{20}
\end{equation}
are the representations on $\Rep{10}$ of the first and second
generators. Tensoring these actions together, one finds that the
invariant subspace consists of $\it two$ copies of the vector-like
pair
\begin{equation}
\big( \Rep{1},\Rep{2}, 3, 0 \big) \,,\quad
\big( \Rep{1},\barRep{2}, -3,  0 \big)
\,.
\label{21}
\end{equation}
That is, there are two Higgs-Higgs conjugate pairs occurring as zero
modes of our vacuum.

Putting these results together, we conclude that the zero mode
spectrum of the observable sector 1) has gauge group $SU(3)_{C} \times
SU(2)_{L} \times U(1)_{Y} \times U(1)_{B-L}$, 2) contains {\it three
  families} of quarks and leptons each with a {\it right-handed
  neutrino}, 3) has {\it two} Higgs-Higgs conjugate pairs and 4) {\it
  contains no exotic fields of any kind}. Additionally, there are 5) a
{\it small number} of uncharged vector bundle moduli. These arise
from the invariant subspace of $H^1\big(\Xt, \Vt \otimes
\Vt^*\big)$ under the action of $\Z_3 \times \Z_3$.

Thus far, we have discussed the vector bundle of the observable
sector. However, the vacuum can contain a stable, holomorphic vector
bundle, $V'$, on $X$ whose structure group is in the $E_8'$ of the
hidden sector. As above, one can construct $V'$ by building stable,
holomorphic vector bundles $\Vt'$ over $\Xt$ which are equivariant
under $\Z_3 \times \Z_3$ using the method of ``bundle extensions''.
$V'$ is then obtained by taking the quotient of $\Vt'$ with $\Z_3
\times \Z_3$. The requirement of anomaly cancellation relates the
observable and hidden sector bundles, imposing the constraint that
\begin{equation}
[\mathcal{W}]= 
c_2\big( T\Xt \big) -
c_2\big( \Vt  \big) -
c_2\big( \Vt' \big) 
\label{22}
\end{equation}
must be an effective class. Here $c_2$ is the second Chern class. In
the strongly coupled heterotic string, $[\mathcal{W}]$ is the class of
the holomorphic curve around which a bulk space five-brane is wrapped.
In the weakly coupled case $[\mathcal{W}]$ must vanish. We have
previously constructed $\Xt$ and $\Vt$ and, hence, can compute
$c_{2}(T\Xt)$ and $c_{2}(\Vt)$. Then eq.~\eqref{22} becomes a
constraint on the hidden sector bundle $\Vt'$. The easiest possibility
is that $\Vt'$ is the trivial bundle. However, in this case, we find
that $[\mathcal{W}]$ is not effective.

The next simplest choice is to take $\Vt'$ to have structure group
\begin{equation}
G'=SU(2)
\label{23}
\end{equation}
in $E_8'$. This spontaneously breaks the hidden sector $E_8'$ symmetry to
\begin{equation}
E_8' \longrightarrow E_7
\,.
\label{24}
\end{equation}
Recall that with respect to $SU(2) \times E_7$ the adjoint
representation of $E_8'$ decomposes as
\begin{equation}
\Rep{248}'=
\big( \Rep{1}, \Rep{133} \big) \oplus 
\big( \Rep{3}, \Rep{1}   \big) \oplus 
\big( \Rep{2}, \Rep{56}  \big)
\,.
\label{25}
\end{equation}
We now require that there be no exotic matter fields in the hidden
sector. This imposes the additional constraint that
\begin{equation}
h^1\left( \Xt, \Vt' \right)=0
\,.
\label{26}
\end{equation}
Finally, the requirement that $\Vt'$ be stable implies the conditions 
\begin{equation}
h^0\left( \Xt,\Vt' \right)=0 \,, \quad 
h^0\left( \Xt,\Vt'^* \right)=0 \,, \quad 
h^0\left( \Xt, \Vt' \otimes \Vt'^* \right)=1
\,.
\label{27}
\end{equation}
It can be shown~\cite{future} that vector bundles $\Vt'$ satisfying
eqs.~\eqref{22},~\eqref{23},~\eqref{26} and~\eqref{27} can be
constructed. For these bundles $[\mathcal{W}]$ is non-vanishing and,
hence, this is a vacuum of the strongly coupled heterotic string. The
five-brane wrapped on a holomorphic curve associated with
$[\mathcal{W}]$ contributes non-Abelian gauge fields, but no matter
fields, to the hidden sector. Following the results in \cite{5brane},
we find that the five-brane gauge group is
\begin{equation}
G'_5=U(6)
\,.
\label{28}
\end{equation}
Moving in the moduli space of the holomorphic curve, this group can be
maximally broken to $U(1)^6$. We conclude that, within the context of
the strongly coupled heterotic string, our observable sector is
consistent with a hidden sector with gauge group $E_7 \times U(6)$ and
{\it no exotic matter}. There are, additionally, a {\it small number}
of uncharged vector bundle moduli that arise from the invariant
subspace of $H^1\big(\Xt, \Vt' \otimes \Vt'^*\big)$ under $\Z_3
\times \Z_3$, as well as some five-brane moduli.

We now exhibit a hidden sector, compatible with our observable sector,
that has no five-branes; that is, for which
\begin{equation}
[\mathcal{W}]=0
\,. 
\label{29}
\end{equation}
This does not occur for structure group $G'=SU(2)$. From the results
in~\cite{redu}, we expect that the appropriate group may be the
product of two non-Abelian groups. The simplest choice is
\begin{equation}
G'=SU(2) \times SU(2)
\,.
\label{30}
\end{equation}
This bundle, which is the sum of two $SU(2)$ factors $\Vt'=\Vt_1'
\oplus \Vt_2'$, spontaneously breaks the hidden sector $E_8'$ gauge
group to
\begin{equation}
E_8' \longrightarrow Spin(12)
\,.
\label{31}
\end{equation}
With respect to $SU(2) \times SU(2) \times Spin(12)$ the adjoint
representation of $E_8'$ decomposes as
\begin{equation}
\Rep{248}'=
\big(\Rep{3}, \Rep{1}, \Rep{1} \big) \oplus 
\big(\Rep{1}, \Rep{3}, \Rep{1} \big) \oplus 
\big(\Rep{1}, \Rep{1}, \Rep{66} \big) \oplus 
\big(\Rep{1}, \Rep{2}, \Rep{32} \big) \oplus 
\big(\Rep{2}, \Rep{1}, \Rep{32} \big) \oplus 
\big(\Rep{2}, \Rep{2}, \Rep{12} \big)
\,.
\label{32}
\end{equation}
The hidden sector will have no exotic matter fields if
\begin{equation}
h^1\left( \Xt, \Vt_1' \right)=0 \,, \quad 
h^1\left( \Xt, \Vt_2' \right)=0 \,,
\label{33}
\end{equation} 
and
\begin{equation}
h^1\left( \Xt, \Vt_1' \otimes \Vt_2' \right)  =0
\,.
\label{34}
\end{equation}
Finally, note that the stability of each bundle $\Vt'_i$,
$i=1,2$ implies the conditions
\begin{equation}
h^0\left( \Xt, \Vt'_i \right)=0 \,, \quad 
h^0\left( \Xt, \Vt'^*_i \right)=0 \,, \quad 
h^0\left( \Xt, \Vt'_{i} \otimes \Vt'^*_{i} \right) =1
\,, \quad i=1,2
\,.
\label{35}
\end{equation}
Subject to eq.~\eqref{22} and the condition eq.~\eqref{29} that there
be no five-brane, we are unable to simultaneously satisfy all of the
constraints in eqs.~\eqref{33},~\eqref{34} and~\eqref{35}. Demanding
that the stability conditions eq.~\eqref{35} hold, it is possible to
choose $\Vt'_i$, $i=1,2$ so that only the first condition
in eq.~\eqref{33} is fulfilled. One finds that, minimally,
\begin{equation}
h^1\left( \Xt, \Vt_2' \right)=4
\label{36}
\end{equation}
and
\begin{equation}
h^1\left( \Xt, \Vt_1' \otimes \Vt_2' \right)=18
\,.
\label{37}
\end{equation}
However, we can show that the $\Z_3 \times \Z_3$ action on
$H^1\big(\Xt, \Vt_2'\big)$ has no invariant subspace. It follows that
the associated matter fields will be projected out under the quotient
by $\Z_3 \times \Z_3$. Unfortunately, this is not the case for
$H^1\big(\Xt, \Vt_1' \otimes \Vt_2'\big)$. Here, we find that the
$\Z_3 \times \Z_3$ action is two copies of its regular representation,
which leaves a two-dimensional subspace of $H^1\big(\Xt, \Vt_1'
\otimes \Vt_2'\big)$ invariant. Hence, after quotienting by $\Z_3
\times \Z_3$, one finds two $\Rep{12}$ multiplets of $Spin(12)$. We
conclude that, for vacua with no five-branes, our observable sector is
consistent with a hidden sector with gauge group $Spin(12)$ and {\it
  two} $\Rep{12}$ {\it multiplets of exotic matter}. There are also
vector bundle moduli arising from the $\Z_3 \times \Z_3$ invariant
subspace of $H^1\big(\Xt, \Vt' \otimes \Vt'^*\big)$. These vacua can
occur in the context of both the weakly and strongly coupled heterotic
string.


\paragraph{Acknowledgments}
We are grateful to P.~Candelas, R.~Donagi, P.~Langacker, 
B.~Nelson, R.~Reinbacher
and D.~Waldram for enlightening discussions. We also thank 
L.~Iba{\~n}ez, H.P.~Nilles, G.~Shiu and A.~Uranga for helpful
conversations. This research was supported in part by the Department
of Physics and the Math/Physics Research Group at the University of
Pennsylvania under cooperative research agreement DE-FG02-95ER40893
with the U.~S.~Department of Energy and an NSF Focused Research Grant
DMS0139799 for ``The Geometry of Superstrings.'' T.~P.~is partially
supported by an NSF grant DMS 0104354 and DMS 0403884.


\end{document}